\documentclass[twocolappendix, apj]{emulateapj}
\usepackage{natbib}
\usepackage{epsfig}
\usepackage{amsmath}
\usepackage{multirow}

\def\hMpc{~h^{-1}{\rm Mpc}}
\def\Msolar{\textrm{M}_{\odot}}

\slugcomment{Submitted 2015 October 7; accepted 2016 May 25}

\shorttitle{Significant BAO from galaxy clusters}
\shortauthors{HONG ET AL.}

\begin{document}

\title{A detection of baryon acoustic oscillations from the distribution of 
  galaxy clusters}

\author{
Tao~Hong, 
J. L.~Han, 
and
Z. L.~Wen 
}
\affil{
National Astronomical Observatories, Chinese Academy
  of Sciences, 20A Datun Road, Chaoyang District, Beijing 100012,
  China. bartonhongtao@gmail.com, hjl@nao.cas.cn}

\begin{abstract}
We calculate the correlation function of 79,091 galaxy clusters in the
redshift region of $z \leq 0.5$ selected from the WH15 cluster
catalog. With a weight of cluster mass, a significant baryon acoustic
oscillation (BAO) peak is detected on the correlation function with a
significance of $3.7 \sigma$. By fitting the correlation function with
a $\Lambda$CDM model curve, we find $D_v(z = 0.331) r_d^{fid}/r_d =
1261.5 \pm 48$~Mpc which is consistent with the Planck 2015
cosmology. We find that the correlation function of the higher mass
sub-sample shows a higher amplitude at small scales of $r < 80\hMpc$,
which is consistent with our previous result. The 2D correlation
function of this large sample of galaxy clusters shows a faint BAO
ring with a significance of $1.8\sigma$, from which we find that the
distance scale parameters on directions across and along the
line-of-sight are $\alpha_{\sigma} = 1.02 \pm 0.06$ and $\alpha_{\pi}
= 0.94 \pm 0.10$, respectively.
\end{abstract}

\keywords{cosmology: observations --- galaxies: clusters: general 
 --- large-scale structure of universe}

\section{Introduction}
\label{sec:intro}
The matter distribution in the Universe is homogeneous and isotropic
on large scales. However, large-scale structures start to emerge from
the matter distribution on smaller scales \citep[$\lesssim
  100\hMpc$,][] {Geller1989, Gott2005, Scrimgeour2012}. The Baryon
Acoustic Oscillations (BAO) are an imprint of the oscillations in the
early Universe when baryons and photons were tightly coupled
\citep{Peebles1970,Sunyaev1970}.  The scale of the BAO can be used as
a `standard ruler' to measure cosmological distances.  For example,
the reduced distance $D_v(z)$ was firstly introduced by
\citet{Eisenstein2005}:
\begin{equation}
\label{eq:Dv}
D_v(z) = \left[ (1+z)^2 D_A(z)^2 \frac{cz}{H(z)}\right]^{1/3},
\end{equation}
\noindent where $H(z)$ is the Hubble parameter and $D_A(z)$ is the
comoving angular diameter distance. The measurement of the BAO
signals provide a powerful tool to constrain the cosmology parameters
which determine $D_v(z)$.

The BAO signal was firstly detected by \citet{Eisenstein2005} and
\citet{Cole2005} using galaxy redshift data from the Sloan Digital Sky
Survey \citep[SDSS, ][]{York2000} and the 2dF Galaxy Redshift Survey
\citep[2dFGRS,][]{Colless2001}. After that, similar measurements of
the BAO were confirmed by later SDSS data releases \citep{Tegmark2006,
  Percival2007, Kazin2010, Anderson2012, Anderson2014} and other
galaxy surveys \citep{Blake2011, Beutler2011}. In addition to
galaxies, the Ly$\alpha$ forests were used as a tracer to search the
BAO signal at higher redshifts. For example, the BAO feature was
detected clearly by using SDSS Ly$\alpha$ forest samples at redshift
$z \sim 2.3$ \citep{Busca2013, Delubac2015}.

Galaxy clusters are the largest gravitationally bound systems in the
Universe, trace the higher density peaks in the matter distribution
field than galaxies, which makes them a great probe for BAO
detection. By calculating the 2-point correlation function and power
spectrum of maxBCG clusters \citep{Koester2007a, Koester2007b},
\citet{Estrada2009} and \citet{Hutsi10} reported weak detections of
BAO signature.  \citet{Hong2012} extracted a spectroscopic sample of
13,904 clusters from \citet{Wen2009} in the redshift region of $z \leq
0.4$, and detected the BAO signature from the cluster correlation
function with a significance of $\sim 1.9
\sigma$. \citet{Veropalumbo2014} further improved this result with a
significance of $\sim 2.5 \sigma$ by using an updated cluster catalog
of \citet{Wen2012}.

In this paper, we calculate and analyze the correlation function of
79,091 clusters from \citet[WH15 hereafter]{Wen2015} with the
spectroscopic redshift information updated to SDSS Data Release 12
\citep[SDSS DR12, ][]{Alam2015}.  The cluster sample is described in
Section~\ref{sec:data}. The method to calculate the two-point
correlation function and the theoretical model to analyze the function
are introduced in Section~\ref{sec:cf_1d}, and we present the
correlation function results for the whole sample and 6 sub-samples in
subsections.  The 2D correlation function of this currently largest
sample of galaxy clusters are presented and discussed in
Section~\ref{sec:cf_2d}.  Conclusions are given in
Section~\ref{sec:discuss}.

\begin{figure*}
\centering
\includegraphics[width=0.4\textwidth, angle=-90]{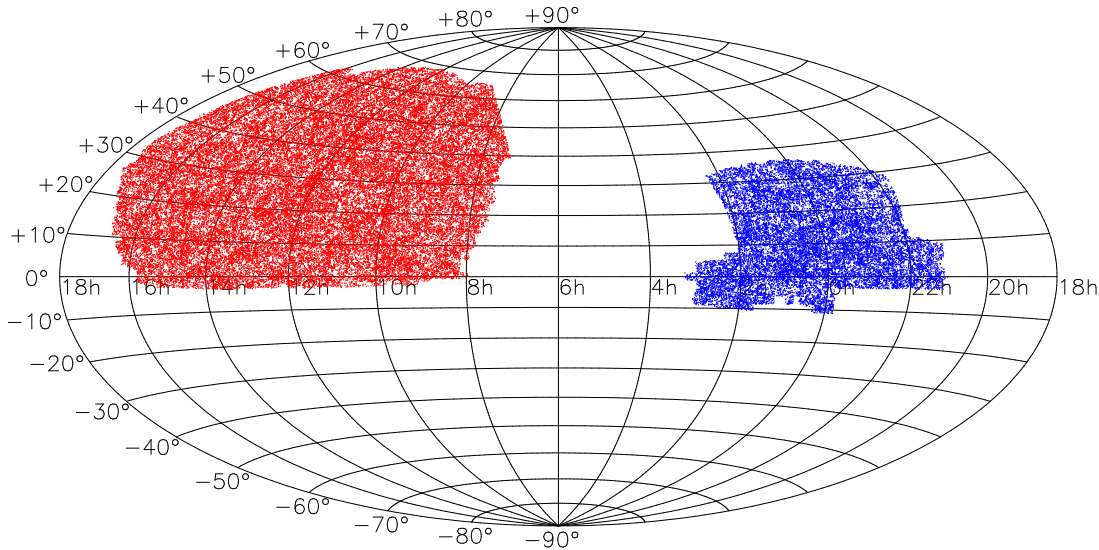}
\caption{Sky distribution of 79,091 clusters in our sample, with an
  Aitoff projection centered at (RA, DEC) $= (6h, 0^\circ)$. There are
  57,647 clusters in the Northern Galactic Cap and 21,444 clusters in
  the Southern Galactic Cap.}
\label{fig:skycover}
\end{figure*}

Throughout this paper, we adopt a flat $\Lambda$CDM cosmology
following Planck 2015 results \citep{Planck2015}, with $h=0.68$,
$\Omega_{m}=0.31$, $\Omega_{\Lambda}=0.69$, $\sigma_{8}=0.81$, where
$h\equiv H_{0}/100~{\rm kms^{-1}Mpc^{-1}}$.

\section{Data}
\label{sec:data}
Using the photometric data from SDSS-III, \citet{Wen2012} identified
132,684 galaxy clusters with a redshift range of $z < 0.8$. All these
clusters have a richness of $R_{L*} \geq 12$ and more than 8 member
galaxies within $r_{200}$. Monte Carlo simulations give a false
detection rate of less than 6\% for the whole catalog. The
completeness is more than 95\% in the redshift range of $z < 0.42$ for
massive clusters with $M_{200} > 1 \times 10^{14} \Msolar$.  By
applying a new richness estimation together with the latest SDSS DR12
spectroscopic data \citep{Alam2015}, WH15 detected 25,000 high
redshift clusters which helps to get a high completeness in the region
of $z < 0.6$ for clusters of $M_{500} > 1 \times 10^{14} \Msolar$.

\begin{figure}
\centering
\includegraphics[width=0.4\textwidth, angle=-90]{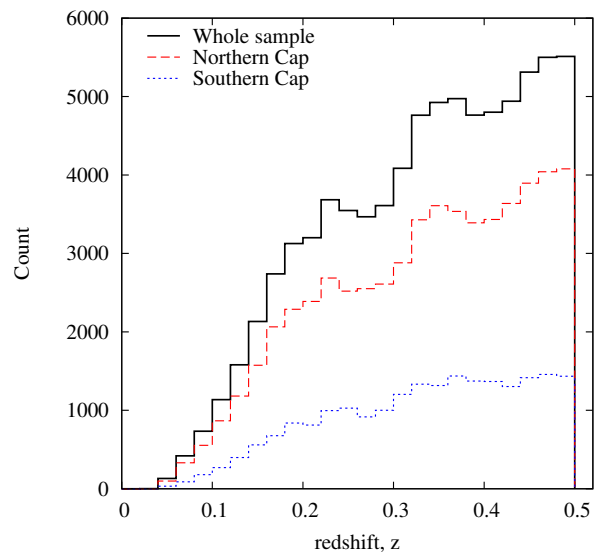}
\caption{Redshift distribution of 79,091 clusters in our sample as
  indicated by black solid line. The dashed line and dotted line
  indicate the distributions of clusters in the Northern Cap and
  Southern Cap, respectively.}
\label{fig:redshift}
\end{figure}

Although the photometric redshift is good enough for identifying the
galaxy clusters, its large uncertainties will affect correlation
function calculations and hence obstruct the detection of BAO
signature \citep{Blake2005,Zhan2008}.  For this work, we use a sample
of 79,091 clusters derived from the WH15 cluster catalog, which have a
spectroscopic redshift from SDSS DR12 data \citep{Alam2015}, including
57,647 clusters from the Northern Galactic Cap and 21,444 clusters
from the Southern Galactic Cap, as shown in Figure~\ref{fig:skycover}.
The whole sample covers a sky region of $\sim 11,000$ square degree in
total. To make sure our sample has a high completeness, we only use
the spectroscopic clusters within the redshift range of $z \leq 0.5$
with a mean redshift $\overline{z} = 0.331$ (see
Figure~\ref{fig:redshift}).

\section{The two-point correlation functions}
\label{sec:cf_1d}
We calculate the 2-point correlation function $\xi(r)$ of cluster
samples using the Landy-Szalay estimator \citep{Landy1993}:
\begin{equation}
\label{eq:LS}
\xi(r)=\left[DD(r)\frac{N_{RR}}{N_{DD}}-2\;DR(r)\frac{N_{RR}}{N_{DR}}+RR(r)\right]/RR(r),
\end{equation}
where $DD(r)$, $DR(r)$ and $RR(r)$ stand for the weighted number of
data-data pairs, data-random pairs and random-random pairs within a
separation annulus of $r\pm\Delta r/2$, respectively.  $N_{DD}$,
$N_{DR}$ and $N_{RR}$ are the weighted normalization factors.  The
random sample used here is 16 times larger than the data sample, which
minimizes the shot noise effect during the calculations. The random
sample shares the same sky area and the same redshift distribution as
the real cluster sample.

\begin{figure}
\centering
\includegraphics[width=0.4\textwidth, angle=-90]{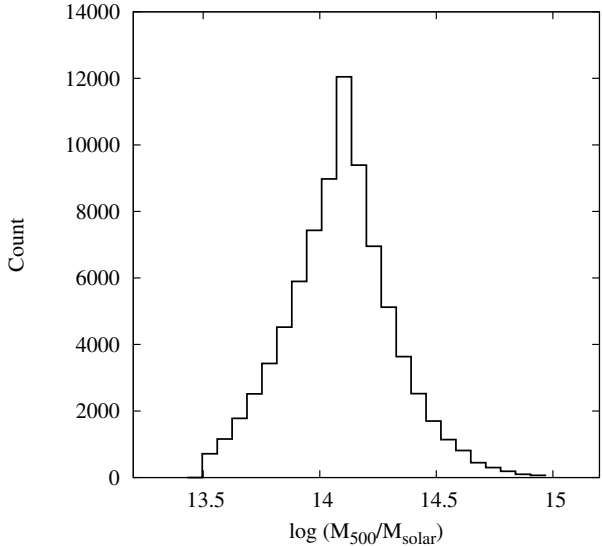}
\caption{Mass distribution of clusters in our cluster sample.}
\label{fig:mass}
\end{figure}

\label{sec:weighting}
More massive galaxy clusters trace more massive dark matter halos,
which should reflect large-scale structures with a larger weight.  To
reveal the BAO feature from the complex matter distribution
background, the more massive clusters should have higher weights than
low mass ones.  The galaxy clusters in this sample have a mass in the
range from $10^{13.5} \Msolar$ to $10^{15} \Msolar$ as shown in
Figure~\ref{fig:mass}.  \citet{Wen2015} have related the cluster mass
with the $r$-band optical luminosity or the richness in their paper.
Here, we adopt a linear weight for cluster mass as
\begin{equation}
\label{eq:weight_mass}
w_{\textrm{mass}}=M_{500}/10^{14}\textrm{M}_{\odot},
\end{equation}
where $M_{500}$ is the cluster mass within the radius where the mean
density is 500 times of the critical density of the Universe (see WH15
for more details).

The completeness of clusters in the sample depends on the mass of
cluster. The completeness can reach 100\% in the high mass end of the
sample distribution, but only about 50\% in the low mass end. To
correct the effect of the detection rate, we apply a weight of
$w_{\textrm{completeness}}$ as the reciprocal of the mass-dependent
detection rate provided in the Figure~6 of \citet{Wen2012}.  The total
weight of the $i^{th}$ cluster for the 2-point correlation function is
thus taken as the combination of the above two weights:
\begin{equation}
\label{eq:weight}
w_i = w_{\textrm{mass}} \times w_{\textrm{completeness}}.
\end{equation}
%
The error covariance of the correlation function is estimated by using
the log-normal mock catalogs. The log-normal error estimation method
was introduced by \citet{Coles1991}, and adopted by several BAO
analysis works \citep[e.g.][]{Beutler2011, Blake2011}. We create the
log-normal realizations using a model power spectrum:
\begin{equation}
\label{eq:model_power_log}
P(k) = b^2(1+\frac{2}{3}\beta+\frac{1}{5}\beta^2)P_{lin}(\overline{z},k),
\end{equation}
where $b$ is the bias measured by fitting the cluster correlation
function using the model curve with the covariance matrix estimated by
the jackknife method, $\beta = \Omega_m^{0.55}/b$,
$P_{lin}(\overline{z},k)$ is the linear power spectrum obtained from
the {\sc camb} package \citep{Lewis2000} at the mean redshift
$\overline{z} = 0.331$.  In total, 100 log-normal mock catalogs are
generated in the boxes of $3000 \times 3000 \times 3000\hMpc$ with
$600 \times 600 \times 600$ cells. The large box size makes sure that
the mock catalogs can cover the whole survey volume of the cluster
catalog, the cell size of $5\hMpc$ is a half of the bin size of our
correlation function measurements. The log-normal mock distribution is
smooth at scales smaller than the cell size. Correlation functions are
calculated for every mock catalog, the covariance matrix is then
generated by:
\begin{equation}
\label{eq:err}
C_{ij}=\frac{1}{N-1}\sum_{k=1}^{N}\left(\xi^{k}_{i}-\overline{\xi}_{i}\right)\left(\xi^{k}_{j}-\overline{\xi}_{j}\right),
\end{equation}
where $N=100$ is the number of mock catalogs, $\xi^{k}_{i}$ is the
correlation function value of the $k^{th}$ mock at the $i^{th}$ bin of
$r$ values, and $\overline{\xi}_{i}$ represents the mean value of the
all 100 mock catalogs at the $i^{th}$ bin. The error bars of $\xi(r)$
are given by the diagonal elements as $\sigma_{i}=\sqrt{C_{ii}}$.

The jackknife error estimation method is adopted for the mass weight
comparison. The details about the jackknife method and the comparison
between log-normal and jackknife covariance matrices are discussed in
the appendix.

We calculate the correlation function and the uncertainty in 18 bins
from $20\hMpc$ to $200\hMpc$.  The analysis are made not only on the
whole sample of 79,091 clusters, but also on six sub-samples divided
according to sky region (Northern Cap and Southern Cap), or the
redshift ranges ($z \leq 0.35$ and $0.35 < z \leq 0.5$) or the
different cluster mass ($\textrm{M}_{500} \leq 1 \times 10^{14}
\textrm{M}_\odot$ and $\textrm{M}_{500} > 1 \times 10^{14}
\textrm{M}_\odot$).

\label{sec:theory}
After that, we analyze the correlation function of galaxy clusters
with a $\chi^2$ fitting to a $\Lambda$CDM model.  First, the linear
matter power spectra $P_{\textrm{lin}}(z, k)$ are computed at each
central value of redshift bin shown in the Figure~\ref{fig:redshift}
using \textsc{camb} package \citep{Lewis2000}. The no-wiggle
approximation of the linear matter power spectrum $P_{\textrm{nw}}(z,
k)$ is generated by fitting the matter power spectrum with the model
described in \citet{Eisenstein1998}. The template power spectrum with
non-linear evolution effects is \citep{Xu2012}
\begin{equation}
\begin{split}
P_{\textrm{template}}(z, k) =& \left(P_{\textrm{lin}}(z, k) -
P_{\textrm{nw}}(z, k)\right)\exp \left(-\frac{k^2
  \Sigma_{\textrm{nl}}^2}{2}\right)\\
  &+P_{\textrm{nw}}(z, k),
\end{split}
\end{equation}
where $\Sigma_{\textrm{nl}}$ is a parameter modeling the non-linear
degradation \citep{Eisenstein2007, Crocce2008, Seo2008, Xu2012}, we
choose $\Sigma_{\textrm{nl}} = 8\hMpc$ in the analysis.  The template
correlation function with damped BAO at each redshift is then given by
\begin{equation}
\xi_{\textrm{template}}(z, r) =\int\frac{k^2
  dk}{2\pi^2}P_{\textrm{template}}(z, k) j_0(kr)
\exp\left(-k^2a^2\right),
\end{equation}
where $j_0(kr)$ is the zeroth-order spherical Bessel function, the
Gaussian term gives a high-$k$ damping during the transformation with
$a = 1\hMpc$, which is significantly smaller than the scale of the
structure we are interested in.  The ``averaged'' template correlation
function $\xi_{\textrm{template}}(r)$ is then generated by weighting
the template correlation functions at each redshift using the
corresponding number counts $n(z)$ in the redshift bins.  Finally, we
fit the cluster correlation function using a model form of
\begin{equation}
\xi_{\textrm{model}}(r) = b^2\xi_{\textrm{template}}(\alpha r)+A(r),
\end{equation}
where
\begin{equation}
A(r)=\frac{a_1}{r^2}+\frac{a_2}{r}+a_3.
\end{equation}
$b^2$, $\alpha$, $a_1$, $a_2$ and $a_3$ are free parameters, $b^2$,
$a_1$, $a_2$ and $a_3$ are marginalized finally.  The $\chi^2$ fitting
runs in the parameter space of $0.80 \leq \alpha \leq 1.20$, where we
fix the other cosmological parameters to the Planck 2015 values of
$\Omega_b= 0.0484$, $n_s = 0.97$, $\sigma_8 = 0.81$, $\Omega_m =
0.31$, $\Omega_\Lambda = 0.69$ and $h = 0.68$. In this fiducial
cosmology, the distance parameter $D_v$ at redshift $z = 0.331$ is
$D_v^{fid} (z = 0.331) = 1301.9~\textrm{Mpc}$.
\label{sec:result}
\begin{table}[h]
\caption[]{BAO fitting results of the cluster sample and sub-samples}
\label{tab:result}
\centering
\begin{tabular}{lccc}
\hline
\hline
Sample & N & $\alpha$  & $\sigma$ \\
 \hline
 \hline
Whole sample & 79091 & 0.969 $\pm$ 0.037 &  3.7\\
\hline
North cap & 57647 & 0.979 $\pm$ 0.058 & 2.2 \\
South cap & 21444 & $\sim 0.939^{\star}$ &  0.7 \\
\hline
$\textrm{M}_{500} > 1 \times 10^{14} \textrm{M}_{\odot}$ & 49207 & 0.979 $\pm$ 0.058 &  2.3 \\
$\textrm{M}_{500} \leq 1 \times 10^{14} \textrm{M}_{\odot}$ & 29884 & $0.960^{\star}$ & 0.6 \\
\hline
$z \leq 0.35$ & 40873 & 0.938 $\pm$ 0.041  & 3.3\\
$0.35 < z \leq 0.50$ & 38218 & 1.020 $\pm$ 0.065 &   2.2\\
\hline
\hline
\end{tabular}
\footnotesize
\tablecomments{we cannot
provide an effective error estimation for the $\alpha$ value for the 
`Southern Cap' and `low mass' sub-samples 
because of very weak signals.}
\end{table}

\subsection{Results of the whole sample}
\begin{figure}
\centering
\includegraphics[width=0.4\textwidth, angle=-90]{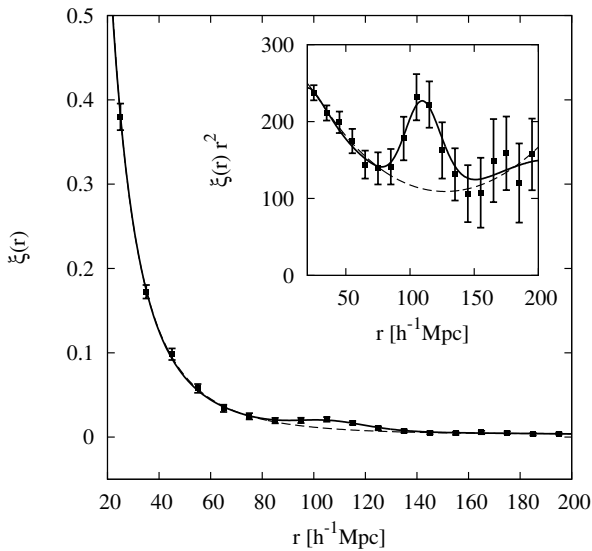}
\caption{Correlation function of 79,091 clusters plotted by black
  squares with error bars. The solid line and dashed line indicate the
  best-fit $\Lambda$CDM model with and without acoustic feature.  In
  the inset $\xi(r) r^2$ is plotted to show the BAO feature more
  clearly. The error bars are estimated via the log-normal method.}
\label{fig:correlation}
\end{figure}
The correlation function of all 79,091 clusters is shown in
Figure~\ref{fig:correlation}. We adopt a weight to correct the
selection bias of the sample and cluster mass. The BAO feature appears
at $r \sim 105\hMpc$ clearly. We do the $\chi^2$ fitting using with
the whole covariance matrix, and find the best-fit $\chi^2 = 6.77$ on
13 degrees of freedom, and the reduced $\chi^2 = 0.52$. A pure CDM
model without the BAO feature is also adopted to fit the correlation
function, which presents a $\chi^2 = 20.29$ and is rejected at $3.7
\sigma$. This is the first time of detecting the BAO signal from a
galaxy cluster sample with a confidence larger than $3 \sigma$. The
best-fit $\Lambda$CDM model offers a constraint on the parameter
$\alpha = 0.969 \pm 0.037$, which gives a constraint on the distance
parameter $D_v$ by $D_v (z=0.331) r_d^{fid}/r_d= 1261.5 \pm
48~\textrm{Mpc}$. See Table~\ref{tab:result} for a summary.

\begin{figure}
\centering
\includegraphics[width=0.4\textwidth, angle=-90]{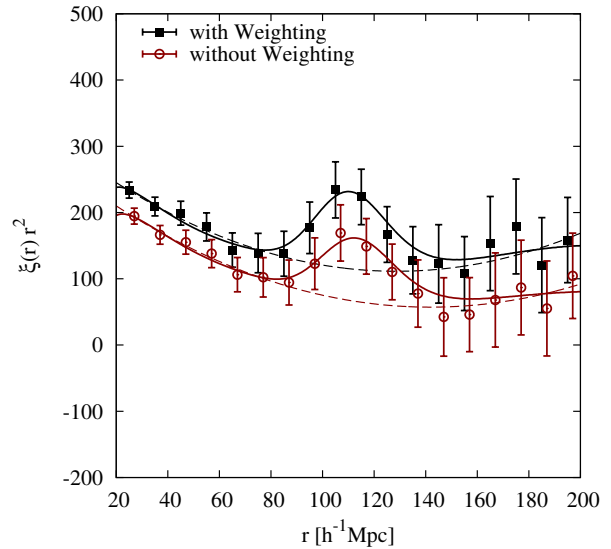}
\caption{The correlation functions of the whole sample with (squares)
  and without (circles, shifted to right by $2\hMpc$ for clarity)
  weights during the calculations. The error bars are estimated by the
  jackknife method. The solid lines and dashed lines indicate the
  best-fit $\Lambda$CDM curves with and without acoustic feature.}
\label{fig:weight}
\end{figure}

\label{sec:compare_weight}
We compare the correlation function of the whole cluster sample
\textrm{without} weighting (i.e. all clusters share the same weight
equals to 1), and compare the result in Figure~\ref{fig:weight}.
Since the log-normal method could not provide the cluster mass for the
mock catalogs, so we use the jackknife method which employs the
original cluster mass of the data catalog in this comparison.  Because
the net effect of the weighting algorithm is giving higher weights to
more massive clusters, the weighting pulls the correlation function up
to the higher amplitude with a detection confidence of $3.9\sigma$,
while the non-weighted calculation gives a confidence of
$3.1\sigma$. We conclude that the mass weight can help the BAO
detection and dose not move the BAO signal position. The best fitted
$\alpha$ value is $\alpha = 0.972$ with the mass weight, compared with
$\alpha = 0.971$ without the mass weight. Therefore, the weightings
are used in all following calculations for sub-samples.

\subsection{Results for two sky regions}
\begin{figure}
\centering
\includegraphics[width=0.4\textwidth, angle=-90]{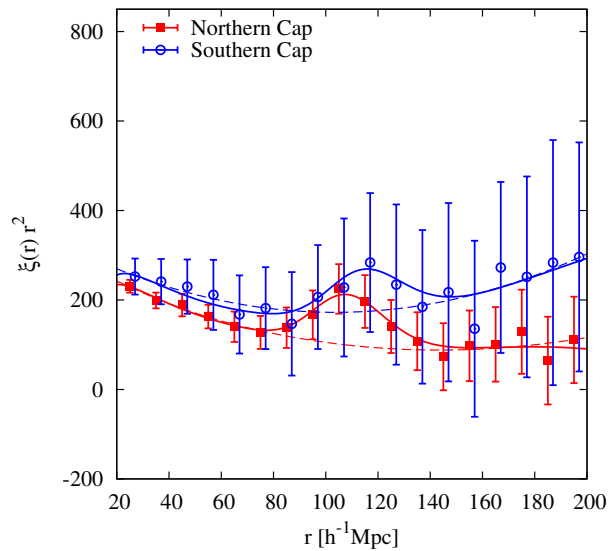}
\caption{The correlation function for BAO detection from galaxy
  clusters in the Northern Cap (squares) and the Southern Cap
  (circles, shifted to right by $2\hMpc$ for clarity), as we do in
  Figure~\ref{fig:correlation}.}
\label{fig:north_south}
\end{figure}
We also calculate the correlation function with the weights for
clusters in the Northern Cap and Southern Cap separately. The
correlation functions are shown in the Figure~\ref{fig:north_south}
with the best-fit model lines. The BAO feature on the Northern Cap is
clear, with a detection confidence of $ 2.2 \sigma$. Due to the
smaller sample size, the BAO signal on the Southern Cap is week, which
has a confidence of $ 0.7 \sigma$.

We notice that the correlation function of the Southern Cap sample has
a higher amplitude comparing with the correlation function of the
Northern Cap sample.  The BAO bump on the correlation function of the
Southern Cap sub-sample also shows a shift towards to the larger scale
direction, the model fitting reports a central value of the distance
parameter of $\alpha = 0.939$, which deviates from the model
prediction, but the low signal-to-noise ratio of the BAO bump leads a
difficulty to estimate the measuring accuracy, the flat $\chi^2$
distribution makes the attempts of determining the $1 \sigma$ error
bar failed.

\subsection{Results for different mass ranges}
\begin{figure}
\centering
\includegraphics[width=0.4\textwidth, angle=-90]{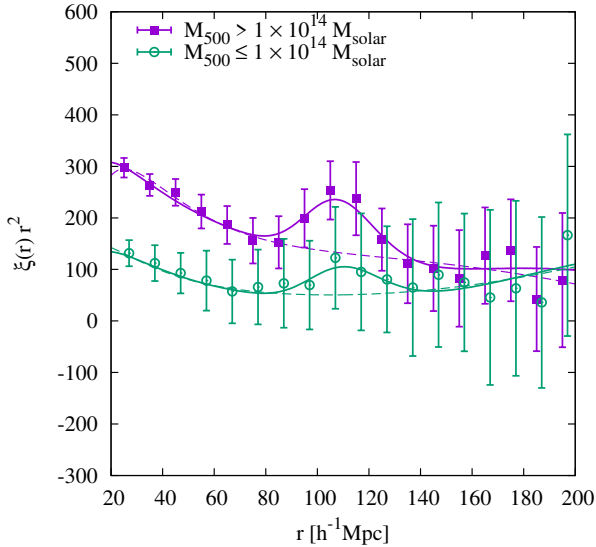}
\caption{The same as Figure~\ref{fig:north_south} but for the high 
mass sample (squares) and the low mass (circles, shifted to right by
  $2\hMpc$ for clarity) clusters.}
\label{fig:mass_bin}
\end{figure}
\begin{figure}
\centering
\includegraphics[width=0.4\textwidth, angle=-90]{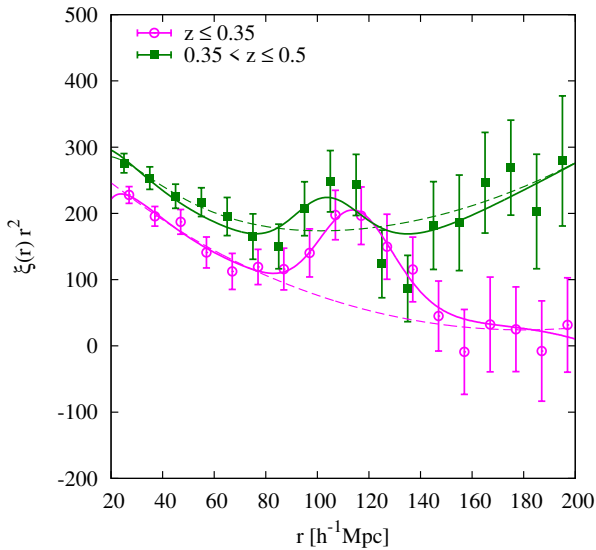}
\caption{The same as Figure~\ref{fig:north_south} but for clusters at 
  high redshift (squares) and low redshift (circles, shifted to right by
  $2\hMpc$ for clarity).}
\label{fig:redshift_bin}
\end{figure}
To compare the correlation function of clusters with different masses,
we make two sub-samples. The high mass sub-sample contains 49,207
clusters with a mass of $\textrm{M}_{500} > 1 \times 10^{14}
\textrm{M}_{\odot}$, the low mass sub-sample has 29,884 clusters with
a mass of $\textrm{M}_{500} \leq 1 \times 10^{14}
\textrm{M}_{\odot}$. The correlation functions of these two
sub-samples are presented in the Figure~\ref{fig:mass_bin}. A clear
BAO signal is detected in the high mass sub-sample with a confidence
of $2.3 \sigma$, while the BAO bump of low mass sub-sample is very
week, only $0.6 \sigma$. Like the `Southern Cap' sub-sample, we cannot
provide $1 \sigma$ error for the low-mass sub-sample because of the
low signal-to-noise ratio.

In small scales of $r < 80\hMpc$, we note the amplitude of correlation
function for high mass clusters is systematically higher than the low
mass ones. \citet{Hong2012} analyzed the correlation functions in
small scales of sub-samples with different cluster richness, found
that the correlation length and then the amplitude of the correlation
function are proportional to the cluster richness. It is expected that
clusters with high masses trace the more massive halos, which leads a
stronger correlation than the low mass sub-sample.  Therefore the
result here is consistent with our previous conclusion.

\subsection{Results for different redshift ranges}

The whole sample is split into two sub-samples by the redshift. The
low redshift sub-sample contains 40,873 clusters with the redshift of
$z \leq 0.35$, the high redshift sub-sample contains 38,218 clusters
in the redshift region of $0.35 < z \leq 0.5$. The correlation
functions of high and low redshift sub-samples are shown in the
Figure~\ref{fig:redshift_bin}. Both of the correlation functions show
BAO signals at the scale of $r \sim 105\hMpc$, the BAO peak detection
confidence is $3.3 \sigma$ and $2.2 \sigma$ on the low and high
redshift sub-samples, respectively.

The correlation amplitude is also found to be different for these two
sub-samples with the scales of $r < 80\hMpc$. The difference is due to
the different cluster mass distributions in the two samples. In the
higher redshift region, luminous and massive galaxies have larger
chances to be spectroscopically observed, which makes our high
redshift sub-sample contains relatively more massive clusters than the
low redshift sample. The mean mass of the high redshift sample is
$M_{500} = 1.42 \times 10^{14} \textrm{M}_{\odot}$, while, the mean
mass of the low redshift sample is $M_{500} = 1.24 \times 10^{14}
\textrm{M}_{\odot}$.

\subsection{Discussions}
\citet{Tojeiro2014} calculated both the correlation function and power
spectrum of 313,780 galaxies from SDSS DR11 over 7,341 square degrees,
in the redshift range of $0.15 < z < 0.43$ with a mean redshift
$\overline{z}= 0.32$. By fitting the BAO feature, they provided a
distance measurement of $D_V(0.32)=1264 \pm 25 (r_d/r_{d,fid})$, with
a measuring accuracy of 1.9\%. In comparison, our cluster sample has a
similar redshift coverage and contains 79,091 clusters, only about
25\% of the galaxy sample size used by \citet{Tojeiro2014}. We detect
the BAO signal by $3.7\sigma$ and get a distance measurement of $D_v
(z=0.331) r_d^{fid}/r_d = 1261.5 \pm 48~\textrm{Mpc}$ with a measuring
accuracy of 3.8\%. This implies a potential economical way to study
the large-scale structures in the future. Spectroscopic observations
are very time consuming especially for faint galaxies.  When doing the
large-scale structure studies using clusters, spectroscopic redshifts
are not necessary for every galaxy. We can identify galaxy clusters
from photometry survey data first, and does the spectroscopic
follow-up for BCGs which are bright and can be easily observed.  A
much smaller sample of clusters can provide a fairly accurate
measurement to the cosmological parameters too. We noticed that after
this paper was submitted, \citet{Veropalumbo2016} calculated the
2-point correlation function using the cluster catalog presented by
\citet{Wen2012} and got a distance measurement consistent with ours.

\section{The 2D correlation function}
\label{sec:cf_2d}
\begin{figure}[t]
\centering
\includegraphics[width=0.4\textwidth, angle=-90]{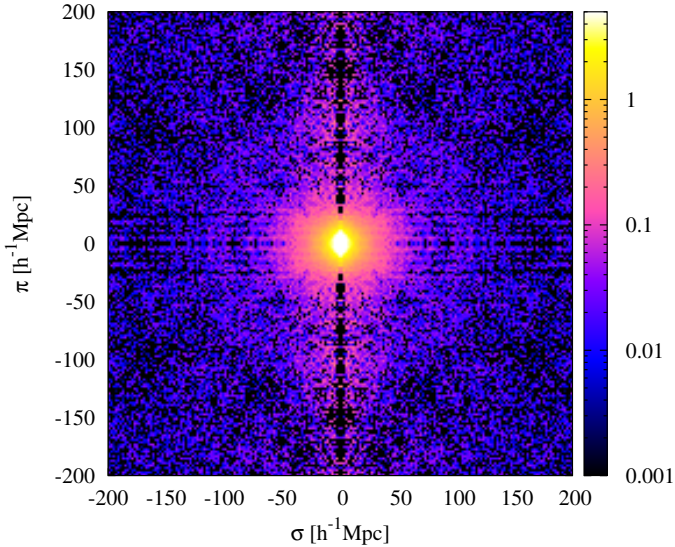}
\caption{The 2D correlation function of the 79,901 clusters. The
  correlation function is binned in 2$\hMpc$ bins.}
\label{fig:2d}
\end{figure}

We calculate the 2D correlation function of the 79,901 clusters
following the same estimator and same weighting method described by
Equation~\ref{eq:LS} and Equation~\ref{eq:weight}.  The result is
shown in Figure~\ref{fig:2d}, where $\pi$ is the separation between
two clusters along the line-of-sight and $\sigma$ is the separation
across the line-of-sight. The faint BAO ring appears at the scale of
$r \sim 105\hMpc$.

\begin{figure}[t]
\centering
\includegraphics[width=0.4\textwidth, angle=-90]{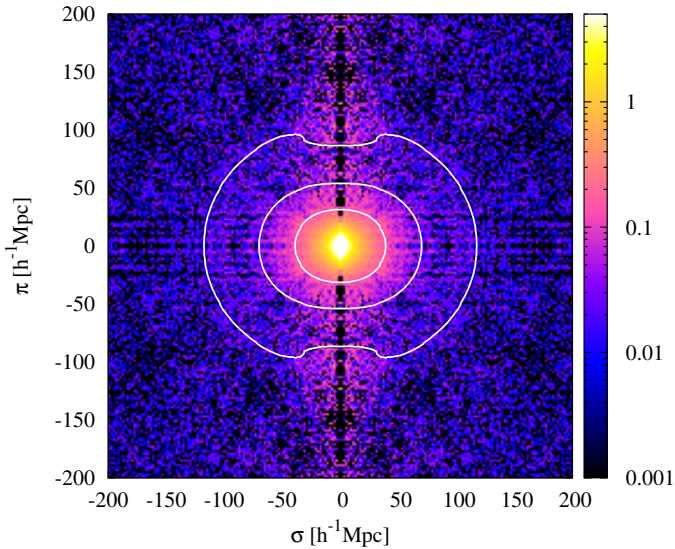}
\caption{2D correlation function with the best-fit theoretical
  correlation function as white contours.}
\label{fig:2d-fitted}
\end{figure}
%
%
Following \citet{Hamilton1992} and \citet{Chuang2013} 
we build a theoretical 2D correlation 
function:
\begin{equation}
\begin{split}
\xi_{\textrm{template}}(\sigma,\pi) =&\xi_0^{\textrm{template}}(r)P_0(\mu)+
\xi_2^{\textrm{template}}(r)P_2(\mu)\\
 &+\xi_4^{\textrm{template}}(r)P_4(\mu),
\end{split}
\end{equation}
with
\begin{equation}
\label{eq:theory-2d_1}
\xi_0^{\textrm{template}}(r)=\left(1+\frac{2\beta}{3}+\frac{\beta^2}{5}\right)\xi(r),
\end{equation}
\begin{equation}
\xi_2^{\textrm{template}}(r)=\left(\frac{4\beta}{3}+\frac{4\beta^2}{7}\right)[\xi(r)-\overline{\xi}(r)],
\end{equation}
\begin{equation}
\xi_4^{\textrm{template}}(r)=\frac{8\beta^2}{35}\left[\xi(r)+\frac{5}{2}\overline{\xi}(r)-\frac{7}{2}\overline{\overline{\xi}}(r)\right],
\end{equation}
where $r=\sqrt{\sigma^2+\pi^2}$, $\mu$ is the cosine of the angle 
between the direction of cluster and the LOS, 
$\beta=\Omega_m^{0.55}/b$. $P_0(\mu)= 1$, 
$P_2(\mu) = \frac{1}{2}\left(3\mu^2-1\right)$ and $P_4(\mu) = \frac{1}{8}
\left(35\mu^4-30\mu^2+3\right)$ are the Legendre polynomials and 
\begin{equation}
\overline{\xi}(r)=\frac{3}{r^3}\int^r_0{\xi(r')r'^2dr'},
\end{equation}
\begin{equation}
\label{eq:theory-2d_2}
\overline{\overline{\xi}}(r)=\frac{5}{r^5}\int^r_0{\xi(r')r'^4dr'},
\end{equation}
where $\xi(r)$ is the theoretical correlation function generated from
the matter power spectrum provided by the \textsc{camb} package using
the same cosmological parameters adopted by the theoretical two-point
correlation function.  Finally, the model correlation function is
given by:
\begin{equation}
\label{eq:theory-2d}
\xi_{\textrm{model}}(\sigma, \pi)=b^2\xi_{\textrm{template}}(\alpha_{\sigma}\sigma, \alpha_{\pi}\pi)+\frac{a_1}{r^2}+\frac{a_2}{r}+a_3,
\end{equation}
where $\alpha_\sigma, \alpha_\pi, a_1, a_2, a_3$ and the bias $b$ are
free parameters, $a_1, a_2, a_3$ and $b$ are marginalized.

By fitting this model to the result in Figure~\ref{fig:2d}, we neglect
the component of `Finger-of-God' \citep{Jackson1972} which arises at
the small scales \citep[e.g.][]{Peacock2001,Ross2007, Beutler2012},
and focus on the feature of BAO ring at the scale range of $40\hMpc
\leq r \leq 150\hMpc$ in the parameter space of $0.80 \leq
\alpha_\sigma \leq 1.20$ and $0.80 \leq \alpha_\pi \leq 1.20$.  We
find the best-fit scale parameters of $\alpha_{\sigma} = 1.02 \pm
0.06$ and $\alpha_{\pi}=0.94 \pm 0.10$, respectively.  By replacing
the theoretical correlation function $\xi(r)$ with a no-wiggle
correlation function $\xi^{nw}(r)$ in the
Equations~\ref{eq:theory-2d_1} to \ref{eq:theory-2d_2}, we build a
no-wiggle 2D correlation function model, and find the difference of
the fitting $\chi^2$ between the models with and without baryon
feature is $\Delta \chi^2 = 3.4$ which provides a BAO ring detection
confidence of $1.8\sigma$. The best-fit model correlation function is
plotted as contours with the cluster correlation function in
Figure~\ref{fig:2d-fitted}.

\section{Conclusions}
\label{sec:discuss}

\label{sec:conclusion}
We build a galaxy cluster sample based on the updated cluster catalog
published by \citet{Wen2015}, which contains 79,901 clusters in the
redshift range of $z \leq 0.5$ with a mean redshift $\overline{z} =
0.331$. All these clusters have spectroscopic redshift measurements
from the SDSS DR12 data \citep{Alam2015}.

We calculate the 2-point correlation function of the cluster sample
with a weight of cluster mass and sample completeness. The weighting
algorithm not only corrects the selection bias introduced by the
cluster identifying process but also enhances the BAO signal on the
final correlation function.  A baryon acoustic peak is detected at the
scale of $r \sim 105\hMpc$, with a detection confidence of $3.7
\sigma$. This is the first time to detect a significant BAO signal
using a galaxy cluster sample. By fitting the observed correlation
function using a $\Lambda$CDM model, we find a constraint of $\alpha =
0.969 \pm 0.037$ and $D_v (z=0.331) r_d^{fid}/r_d = 1261.5 \pm
48~\textrm{Mpc}$, which show a great consistency with the fiducial
cosmology obtained by the Planck 2015 data.
  
We also calculate the 2D correlation function of the cluster
sample. The faint BAO ring emerges at the scale of $r \sim
105\hMpc$. By fitting the correlation function using a theoretical 2D
correlation function, we detect the BAO ring with a detection
confidence of $1.8\sigma$. Though it is not good enough to detect the
BAO feature in the separated two directions, we get the constraint on
the distance parameters of $\alpha_{\sigma} = 1.02 \pm 0.06$ and
$\alpha_{\pi} = 0.94 \pm 0.10$.

We conclude that the BAO detection via spectroscopically observed BCGs
can easy the survey job, because one can find galaxy clusters first
via photometric data, and then do spectroscopic observations for a
much smaller sample of galaxies.

\begin{acknowledgments}
We thank X. Y. Gao, G. B. Zhao, Y. T. Wang and F. Beutler for useful
discussions and comments. The authors are supported by the National
Natural Science Foundation of China (11473034). TH and ZLW are also
supported by the Young Researcher Grant of National Astronomical
Observatories, Chinese Academy of Sciences.
\end{acknowledgments}
\begin{appendix}
\section{A comparison of covariance matrices estimated by the log-normal and jackknife methods}
The jackknife method estimates the covariance matrix by making
sub-samples based on the original data catalog (internal estimate),
which is somehow different from the external estimate based on
\textit{N}-body simulations or the log-normal realizations. By
comparing the covariance matrix estimated of the internal and external
methods on the scales of $0.1 - 40\hMpc$, \citet{Norberg2009} found
the jackknife method overestimates the variance on small scales of
$\lesssim 2-3\hMpc$, but it works fine on larger scales of $\gtrsim
10\hMpc$. On the BAO scales of $\sim 100\hMpc$, \citet{Beutler2011}
concluded that the jackknife error is noisier and larger than the
log-normal error for the 6dFGS galaxy sample.  Here we compare the
covariance matrices of the correlation function estimated by the
log-normal and jackknife methods.
\begin{figure}[!ht]
\centering
\includegraphics[width=0.4\textwidth, angle=-90]{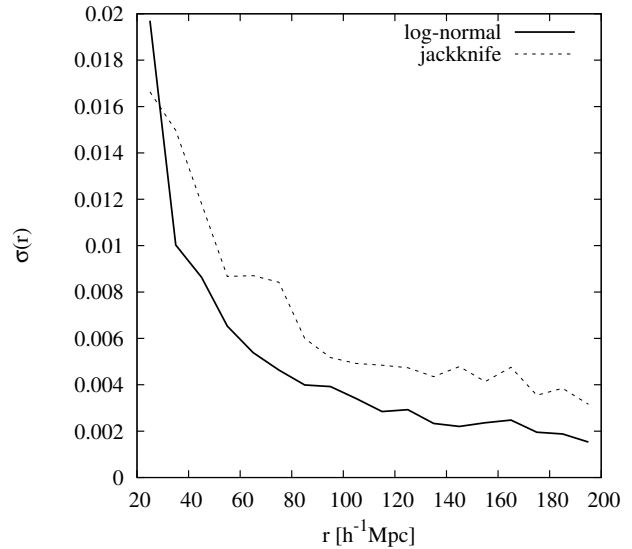}
\caption{The correlation function errors estimated by the log-normal
  method (solid line) and the jackknife method (dashed line)
  respectively.}
\label{fig:log-jack}
\end{figure}
\begin{figure}[t]
\centering
\includegraphics[width=0.4\textwidth, angle=-90]{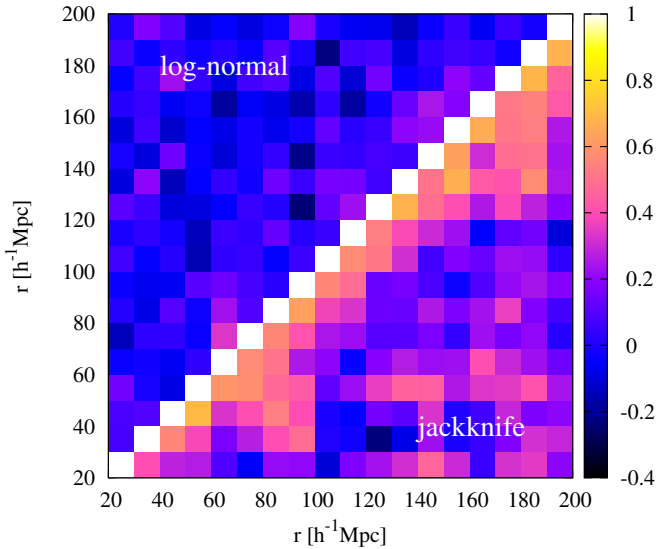}
\caption{The correlation matrix of the log-normal errors and jackknife
  errors.}
\label{fig:log-jack-full}
\end{figure}

We obtain the jackknife covariance matrix by dividing the sky area
into 32 disjoint sub-regions, each sub-region has approximately the
same area with others. The jackknife method is found to be robust when
changing the number of jackknife sub-samples
\citep{Veropalumbo2016}. The 32 jackknife sub-samples are built by
removing the clusters in one sub-region, ensuring that each sub-region
is removed in one sub-sample only. The correlation function is
calculated for each sub-sample following Equation~\ref{eq:LS}. The
covariance matrix is then built up as:
\begin{equation}
\label{eq:err_jack}
C_{ij}=\frac{N-1}{N}\sum_{k=1}^{N}\left(\xi^{k}_{i}-\overline{\xi}_{i}\right)\left(\xi^{k}_{j}-\overline{\xi}_{j}\right),
\end{equation}
where $N=32$ is the number of sub-samples, $\xi^{k}_{i}$ is the
correlation function value of the $k^{th}$ sub-sample at the $i^{th}$
bin of $r$ values, and $\overline{\xi}_{i}$ represents the mean value
of the all 32 sub-samples at the $i^{th}$ bin.

We show the jackknife error in Figure~\ref{fig:log-jack} together with
the log-normal error as a comparison. The jackknife error is found to
be larger than the log-normal error in most of the bins, it is also
noisier than the log-normal error.  Besides the diagonal term of the
covariance matrix, we also show the full matrix in
Figure~\ref{fig:log-jack-full}. The covariance estimated by the
log-normal method is much smoother than the one estimated by the
jackknife method. The elements plotted in
Figure~\ref{fig:log-jack-full} are defined as:
\begin{equation}
r_{ij}=\frac{C_{ij}}{\sqrt{C_{ii}C_{jj}}},
\end{equation}
where $C$ is the covariance matrix.

We noticed that \citet{Veropalumbo2016} also compared the covariance
matrices of the jackknife method and the log-normal method using a
galaxy cluster sample with SDSS III spectroscopic redshift, and they
found a similar conclusion as ours.

\end{appendix}

\bibliographystyle{apj}
\bibliography{bibfile}
\end{document}